\documentclass[amsmath, amssymb, twocolumn, nofootinbib, superscriptaddress, showkeys, aps, prb, 10pt]{revtex4-2}

\usepackage{header}

\begin{document}

\title{Density Saturation and Autocorrelation Characteristics of Interlayer Exciton Ensembles in MoSe$_2$/WSe$_2$ Heterostructures}
\author{Mirco Troue}
\affiliation{Walter Schottky Institute and TUM School of Natural Sciences, TU Munich, Am Coulombwall 4a, 85748 Garching, Germany}
\affiliation{Munich Center for Quantum Science and Technology (MCQST), Schellingstr.~4, 80799 Munich, Germany}
\author{Johannes Figueiredo}
\affiliation{Walter Schottky Institute and TUM School of Natural Sciences, TU Munich, Am Coulombwall 4a, 85748 Garching, Germany}
\affiliation{Munich Center for Quantum Science and Technology (MCQST), Schellingstr.~4, 80799 Munich, Germany}
\author{Alina Scheuermann}
\affiliation{Walter Schottky Institute and TUM School of Natural Sciences, TU Munich, Am Coulombwall 4a, 85748 Garching, Germany}
\affiliation{Munich Center for Quantum Science and Technology (MCQST), Schellingstr.~4, 80799 Munich, Germany}
\author{Sebastian Loy}
\affiliation{Walter Schottky Institute and TUM School of Natural Sciences, TU Munich, Am Coulombwall 4a, 85748 Garching, Germany}
\affiliation{Munich Center for Quantum Science and Technology (MCQST), Schellingstr.~4, 80799 Munich, Germany}
\author{Jonas Kiemle}
\affiliation{Walter Schottky Institute and TUM School of Natural Sciences, TU Munich, Am Coulombwall 4a, 85748 Garching, Germany}
\affiliation{Munich Center for Quantum Science and Technology (MCQST), Schellingstr.~4, 80799 Munich, Germany}
\author{Takashi Taniguchi}
\affiliation{Research Center for Materials Nanoarchitectonics, National Institute for Materials Science, 1-1 Namiki, Tsukuba 305-0044, Japan}
\author{Kenji Watanabe}
\affiliation{Research Center for Electronic and Optical Materials, National Institute for Materials Science, 1-1 Namiki, Tsukuba 305-0044, Japan}
\author{Ursula Wurstbauer}
\affiliation{Institute of Physics, University of M\"unster, Wilhelm-Klemm-Str.~10, 48149 M\"unster, Germany}
\author{Alexander W. Holleitner}
\email{holleitner@wsi.tum.de}
\affiliation{Walter Schottky Institute and TUM School of Natural Sciences, TU Munich, Am Coulombwall 4a, 85748 Garching, Germany}
\affiliation{Munich Center for Quantum Science and Technology (MCQST), Schellingstr.~4, 80799 Munich, Germany}

\date{October 8, 2026}

\begin{abstract}
We investigate interlayer excitons in MoSe$_2$/WSe$_2$ heterostructures using time-resolved photoluminescence and second-order autocorrelation measurements across a wide range of excitation powers and bath temperatures. By analyzing the photoluminescence amplitude and lifetime of the interlayer exciton state with the lowest emission energy, we extract a corresponding density-temperature diagram. The extracted exciton density exhibits a saturation at approximately $10^{13}$~cm$^{-2}$, consistent with values predicted for a Mott transition of interlayer exciton ensembles in van der Waals heterostructures. Moreover, possible interpretations beyond the Mott transition are discussed. We further test whether the density limit is accompanied by a departure from classical photon statistics. The second-order autocorrelation measurements reveal that the photoluminescence maintains an emission characteristic of $g^{(2)}(0) \approx 1$ from low exciton densities up to the maximum reachable density in the investigated exciton ensembles consistent with the lowest-energy exciton ensemble obeying Poissonian statistics.
\end{abstract}


\maketitle

\section{Introduction}\label{sec:intro}

In recent years, the discovery of two-dimensional transition-metal dichalcogenides has opened new routes for exploring excitonic phenomena in reduced dimensional systems. Particularly, interlayer excitons (IXs) formed across the interface of a MoSe$_2$/WSe$_2$ van-der-Waals heterostack exhibit unique optical properties \cite{BrotonsGisbert2024}, including a large exciton binding energy of several hundred meV, a permanent out-of-plane dipole moment, long diffusion lengths, and a long photoluminescence lifetime on the order of 100 ns \cite{Fogler2014,Rivera2015}. The long lifetime suggests that IXs can fully thermalize to an excitonic ground state for sufficiently long detection times, which makes them attractive for the exploration of many-body phenomena at sufficiently high interlayer exciton densities and low enough temperatures \cite{Fogler2014,Rivera2015,Sigl2020,Troue2023,Katzer2023,Cutshall2025}. The density determines the effective strength of interactions between individual IXs and is an important parameter in the buildup of collective behavior \cite{Fogler2014}. At low exciton densities, the interactions are weak and the overall exciton system behaves like a dilute gas of independent particles \cite{Laikhtman2009}. As the density increases, exciton-exciton and exciton-charge carrier interactions can become significant, which eventually leads to spectral broadening and shifts of the photoluminescence characteristics and a modification of the ensemble lifetime \cite{Garrity2024,Blundo2024}. At sufficiently high exciton densities, exciton ensembles are expected to undergo a transition to a correlated exciton phase, where a macroscopic coherence builds up first in individual puddles and then across the full exciton ensemble at low enough temperatures \cite{Fogler2014,Troue2023,Cutshall2025}.

The creation of thermalized high-density ensembles in optically pumped systems remains challenging. Effects such as exciton-exciton repulsion and temperature-driven diffusion locally redistribute the excitons into regions of lower density. To overcome these limitations, various approaches, including electrostatic trapping via tailored gate geometries \cite{Gartner2007,Schinner2011,Shanks2021,Joe2024}, strain engineering \cite{Kremser2020}, and non-Gaussian optical excitation schemes \cite{Troue2026} have been pursued to modulate the overall excitonic potential landscape and to locally increase the exciton density of thermalized excitons. Still, the Mott transition marks the natural density limit of exciton ensembles, since it creates a balance between bound and free charge carriers \cite{Mott1968,Zimmermann1988,Reinholz2002,Nikolaev2004,Stern2008,Snoke2008,Manzke2011,Asano2014,Kirsanske2016}. Close to the Mott transition, small changes in carrier density can result in significant changes in the excitonic properties, such as a change in the binding energy and the absorption characteristics \cite{Steinhoff2017}. For IXs, the Mott-transition has so far been reported in terms of spectral broadening of the photoluminescence \cite{Joe2024,Wang2019,Wietek2024,BrotonsGisbert2021}, an significant increase of the IX-related diffusivity and photoconductance \cite{Wietek2024,Wang2021}, as well as a reduction of the photoluminescence lifetime \cite{Wang2019}. However, whether the Mott transition is the only mechanism limiting the achievable exciton density in optically pumped ensembles, as opposed to, e.g., filling of localized and moir{\'e} states, as well as a photo-doping induced renormalization of moir{\'e}- and disorder-induced modulations of the potential landscape, effectively impacting exciton ensembles, remains an open question. Moreover, a clear connection between the density limit of an IX ensemble and its photon emission statistics has not been demonstrated yet: does the saturated, high-density state still emit photons independently in a Poissonian statistics, or does the onset of a possible Mott transition or a correlated many-body phase leave a signature in the second-order correlation function $g^{(2)}(\tau)$?

Here, we present time-resolved photoluminescence measurements of IXs in MoSe$_2$/WSe$_2$ heterostructures for a large range of excitation powers and bath temperatures. We model the decay dynamics of the observed IX-related photoluminescence by extracting the amplitude and lifetime of the time-resolved traces to estimate the exciton densities for a given temperature. We observe that the exciton density of the lowest energy peak saturates and that the maximum exciton density on two different samples equals $(2.4 \pm 0.1)\times 10^{13}$~cm$^{-2}$ and $(5.6 \pm 0.7)\times 10^{12}$~cm$^{-2}$, respectively. The values are consistent with a predicted Mott transition for IXs in MoSe$_2$/WSe$_2$ heterostructures \cite{Fogler2014}. Moreover, we investigate the second-order autocorrelation function $g^{(2)}(\tau)$ of the excitonic photoluminescence, with $\tau$ the time delay between two photon emission processes. We find $g^{(2)}(0) \approx 1$ within the given time resolution for the energetically lowest interlayer exciton state across the investigated density-temperature diagram. This Poissonian emission behavior persists throughout the investigated temperature range, namely from 1.7K up to 12K. Our results suggest that the Poissonian emission behavior prevails across the whole phase-diagram of interlayer excitons, ranging from an ideal gas of excitons at high temperatures to coherent, degenerate many-body ensembles at low temperature and high densities \cite{Fogler2014,Sigl2020,Troue2023,Katzer2023,Cutshall2025}.

\begin{figure}
  \centering
  \includegraphics{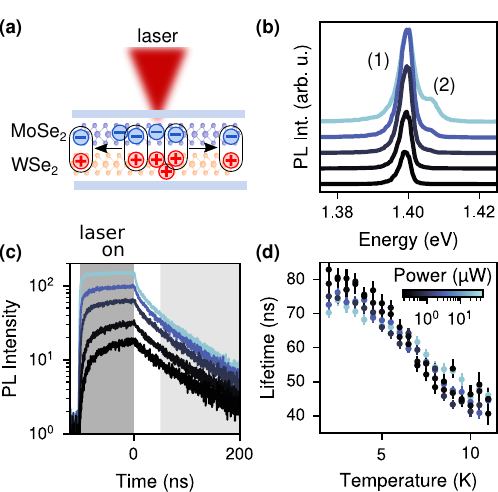}
  \caption{(a) (a) Schematic of the investigated MoSe$_2$/WSe$_2$ H-type heterostructures encapsulated in hBN (blue layers). The excitation laser excites interlayer excitons that are able to expand to the outside of the SM collection spot, as indicated by the black arrows. (b) Spectrally resolved photoluminescence (PL) for a laser power of 0.2 \textmu W, 0.5 \textmu W, 2.3 \textmu W, 7.6 \textmu W and 46.0 \textmu W at 1.7 K under continuous excitation. Spectra are normalized to integration time and vertically offset for clarity. (c) Time-resolved photoluminescence histograms. The laser is switched off at 0 ns. At late times, a single-exponential decay prevails (highlighted in light gray), which we assign to peak (1). (d) PL lifetime extracted from the time-resolved photoluminescence histograms for a time delay between 50 ns and 200 ns vs.\ excitation power and temperature. Colors in (b), (c), and (d) describe equal laser powers.}
  \label{fig:1}
  \phantomsubcaption\label{fig:1:a}
  \phantomsubcaption\label{fig:1:b}
  \phantomsubcaption\label{fig:1:c}
  \phantomsubcaption\label{fig:1:d}
\end{figure}

\section{Results}\label{sec:results}

We investigate optically generated IXs formed across the MoSe$_2$/WSe$_2$ interface in h-BN encapsulated nominally H-type heterostructures, such that the excitons comprise an electron and a hole, whose wavefunction centers are spatially separated toward the MoSe$_2$ and WSe$_2$ layers [\cref{fig:1:a}]. The photoluminescence is collected and analyzed with the help of optical fibers in combination with a spectrometer (see \cref{sec:methods} and Supplemental Information). The photoluminescence features a narrow peak (1) with an energy around 1.4eV at low excitation powers [\cref{fig:1:b}]. At higher powers, a second peak (2) starts to emerge at an energy above the one of peak (1). We interpret both peaks to arise from the recombination of IXs from different sub-ensembles \cite{BrotonsGisbert2024,Fogler2014,Rivera2015,Sigl2020,Troue2023,Cutshall2025,Miller2017,Nayak2017}; with the lowest lying photoluminescence peak usually being assigned to a spin-triplet interlayer state \cite{Yu2018,Gillen2021} in combination with a spatially varying extension of the center-of-mass wavefunction \cite{Richter2024,Figueiredo2025a}. To further investigate the nature of this peak with the lowest energy, we consider time-resolved intensity measurements of the photon emission.

\cref{fig:1:c} depicts the time-resolved photoluminescence traces as a function of excitation power. We utilize optical filters such that only photons with an energy between 1.30 eV and 1.53 eV are detected. While the laser is on, the emission intensity increases monotonically with laser power [dark gray area in \cref{fig:1:c}]. For a time delay between 50 ns and 200 ns [light gray area in \cref{fig:1:c}], a single-exponential decay with a decay time of 40 ns to 80 ns is observed for all investigated excitation powers. We assign the corresponding lifetime to peak (1) as in \cref{fig:1:b}, since this photoluminescence peak occurs for all investigated excitation powers and at the lowest energy. For higher laser powers and at short time delays after the laser is switched off, a second much shorter decay time on the order of 1-2 ns appears, which we assign to peak (2) \cite{Figueiredo2025b}. We note that the colors in \Cref{fig:1:b,fig:1:c} highlight identical laser powers. \Cref{fig:1:d} depicts the photoluminescence lifetimes extracted at a time delay between 50 ns and 200 ns as a function of temperature for several excitation powers. The maximum lifetime of about 80 ns occurs at the lowest temperatures and excitation powers (black data), and it drops by about 50\% over the investigated temperature range. Generally, thermal broadening and phonon-assisted processes are minimal at low power and low temperature \cite{Garrity2024,Moody2015}. As the temperature increases, phonon-assisted recombination and scattering processes of the interlayer excitons are enhanced, reducing the apparent photoluminescence lifetime of the excitons, which qualitatively explains the dependences in \Cref{fig:1:d}.

A systematic estimation of the exciton density at the position of the photoluminescence detection requires the consideration of IXs that propagate out of the collection spot due to diffusion and drift [see arrows in \cref{fig:1:a}], both of which strongly depend on the exciton density \cite{Wietek2024,Gartner2006}. The underlying reasoning is that excitons that move out of the detection spot before recombination reduce the photoluminescence amplitude and therefore, the apparent lifetime as measured within the detection spot. In turn, we investigate the IX-related photoluminescence from two differently-sized detection areas using a single-mode (SM) and a multi-mode (MM) fiber, while in both cases, the excitation occurs via the same single-mode fiber (cf.\ Figs.~S1, S2, and S3). The collection with the multi-mode fiber results in a 440-fold increase in detection area to $A_\mathrm{MM} = 210.2$~\textmu m$^2$ compared to the area of the SM-fiber with $A_\mathrm{SM} = 0.48$~\textmu m$^2$. Fitting the photoluminescence time-traces at a delay time of 50 ns and later in both cases (SM and MM) yields again the apparent lifetimes, but also the extrapolated initial amplitudes of the photoluminescence at the time when the laser is switched off, as in \cref{fig:1:c} (cf.\ Fig.~S4).

\begin{figure}
  \centering
  \includegraphics{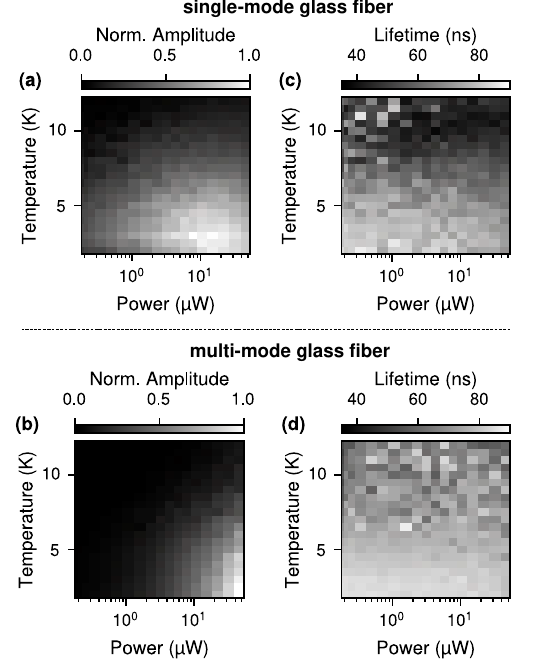}
  \caption{(a) Normalized amplitude of time-resolved PL histograms measured with a single-mode glass fiber and a collection area of $A_\mathrm{SM} = 0.48$~\textmu m$^2$. (b Normalized amplitude of time-resolved PL histograms for a multi-mode fiber detection and a collection area of $A_\mathrm{MM} = 210.2$~\textmu m$^2$. (c) and (d) Corresponding PL lifetimes.}
  \label{fig:2}
  \phantomsubcaption\label{fig:2:a}
  \phantomsubcaption\label{fig:2:b}
  \phantomsubcaption\label{fig:2:c}
  \phantomsubcaption\label{fig:2:d}
\end{figure}

\Cref{fig:2:a,fig:2:b} show these amplitudes of peak (1) normalized to the maximum amplitude for SM- and MM-fiber detections. Generally, both measurements (SM and MM) show a decrease in amplitude with increasing temperature for a given excitation power, as can be explained by interactions with phonons and charge carriers. The most notable difference between the two measurements is the shift of the power at which the maximum amplitude is detected. For the SM-fiber, it is 14.5 \textmu W, while it is 46 \textmu W in the case of the MM-fiber. In our understanding, this observation reflects the additional collection of photoluminescence stemming from excitons that propagate outside of the collection area of the SM-fiber. We note that the maximum excitation power in \cref{fig:2:a,fig:2:b} reflects the situation where peak (2) starts to emerge in the case of the SM-detection [cf.\ \cref{fig:1:b}]. Interestingly, there is a slight decrease of the extracted amplitude of peak (1) for the highest powers in \cref{fig:2:a}. Tentatively, we explain this decrease by a combination of an occupation saturation of the exciton state underlying peak (1), and the presence of free charge carriers as will be discussed below. \cref{fig:2:c,fig:2:d} depict the extracted photoluminescence lifetimes for both measurements. We observe for both that an increased temperature and/or excitation power leads to a reduction in lifetime, as interpreted above in connection with \cref{fig:1:c}. Below 6 K, the observed lifetimes for the SM-fiber detection are comparable to those for the detection with the MM-fiber. Above 6 K, however, the detected lifetimes in the SM case drop more significantly than the ones in the MM case. We explain this observation by a temperature-driven propagation of excitons outside of the detection area of the single-mode glass fiber. For comparison, we estimate a lower bound of the exciton diffusion length from spatial cw-photoluminescence scans to be $\approx$ 0.45 \textmu m at 1.7 K for sample 1 (0.51 \textmu m for sample 2), which explains why a significant fraction of the excitons propagate outside of the excitation spot and the detection area of the single-mode fiber, especially at the highest investigated temperatures. In turn, the detection with the multi-mode glass fiber is nearly unaffected by the relatively small amount of IXs propagating outside of the detection area and captures the photoluminescence lifetime of the exciton ensembles, including diffusive excitons, as excited by the single-mode glass fiber.

\begin{figure}
  \centering
  \includegraphics{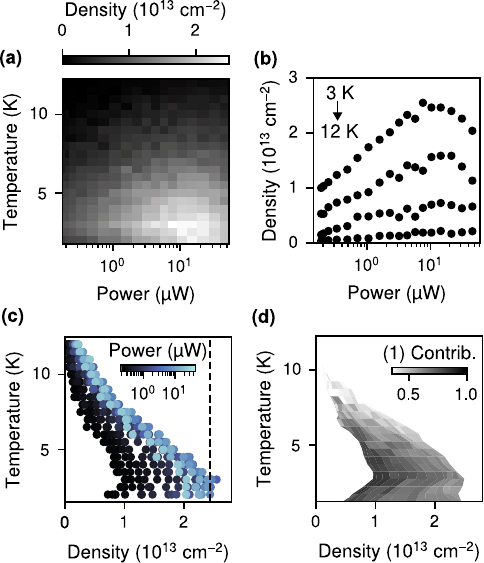}
  \caption{(a) Exciton density within the collection spot for varying excitation powers and temperatures as deduced from peak (1). (b) Extracted temperature line cuts from (a) for 3 K, 6 K, 9 K, and 12 K. (c) Density-temperature line cuts from (a) plotted vs.\ exciton density. The color of the data refers to the excitation power as given in \cref{fig:1:b}. Dashed line highlights the maximum density as reached at the lowest temperature. (d) Extrapolated emission contribution of peak (1) to the total photon emission count at $\tau = 0$~ns [laser off as indicate in \cref{fig:1:c}].}
  \label{fig:3}
  \phantomsubcaption\label{fig:3:a}
  \phantomsubcaption\label{fig:3:b}
  \phantomsubcaption\label{fig:3:c}
  \phantomsubcaption\label{fig:3:d}
\end{figure}

In the following, we estimate the IX density within the collection spot of the single-mode glass fiber during the laser-on period, as it is typically investigated in optical steady-state measurements using a confocal detection scheme. For this, the amplitude of peak (1) from the single-mode measurement, as in \cref{fig:2:a}, is calculated with the lifetime from the multi-mode measurement, as in \cref{fig:2:d} (\cref{sec:methods}, Fig.~S3). In this way, we account for the diffusion and drift losses of excitons propagating to areas outside of the collection spot of the single-mode glass fiber. \Cref{fig:3:a} displays the corresponding exciton density of peak (1) as a function of excitation power and temperature. Below 3.5 K, the calculated density saturates at a maximum value of $(2.4 \pm 0.1)\times 10^{13}$~cm$^{-2}$ for an excitation power of 14.5\textmu W (for sample 2, the saturation density equals $(5.6 \pm 0.7)\times 10^{12}$~cm$^{-2}$, see Fig.~S5). The values are consistent with theoretical \cite{Fogler2014,Steinhoff2017} and experimental \cite{Joe2024,Wang2019,BrotonsGisbert2021,Wang2021} results for a Mott transition of interlayer exciton ensembles in MoSe$_2$/WSe$_2$ heterostructures. Consistently, we observe that the spectral linewidth of peak (1) significantly broadens for laser powers higher than 14.5\textmu W (cf.\ Fig.~S7) and that the extracted photoluminescence amplitude is slightly reduced [cf.\ \cref{fig:2:a}]. Moreover, we observe that the extracted exciton density of the peak (1) sub-ensemble starts to decrease as the pump power increases beyond 14.5\textmu W. The line cuts in \cref{fig:3:b} highlight the evolution of this reversal at four exemplary temperatures, as will be discussed below. \Cref{fig:3:c} depicts the exciton density as deduced from the emission of peak (1) in a density-temperature diagram. The maximum density is reached at the lowest temperature for 14.5 \textmu W (dashed line).

In a next step, we estimate the contribution of peak (1) to the overall photoluminescence by comparing the extrapolated amplitude of peak (1), when the laser is switched off, to the total intensity of the photon emission (see Supporting Information). \Cref{fig:3:d} shows this relative emission contribution of peak (1) in a density-temperature diagram up to the maximum density achievable. The highest contribution from peak (1) is observed at a density of $\approx 1 \times 10^{13}$~cm$^{-2}$ at the lowest temperature (black color), where the contribution of thermal fluctuation and scattering processes (exciton-exciton and exciton-phonon) to a state's photoluminescence are minimized \cite{Moody2015}. We find a maximum contribution of $(87 \pm 3.6)$~\%. In comparison, the contribution of peak (1) to the overall photoluminescence as calculated from the multi-mode measurement exceeds $(95 \pm 2.0)$~\% (cf.\ Fig.~S6). On the one hand, the high values close to unity indicate that the majority of photons are emitted via peak (1) at low temperature [see also \cref{fig:1:b}], suggesting that peak (1) represents the energetically lowest state of the exciton ensemble. On the other hand, the difference between both values represents the IXs propagating outside of the single-mode collection spot. Interestingly, the contribution of peak (1) to the overall photoluminescence reduces for densities above $1\times 10^{13}$~cm$^{-2}$, indicating that, at higher densities, more and more photon emission occurs via peak (2), consistent with the continuous-wave photoluminescence spectra at higher excitation powers in \cref{fig:1:b}.

\begin{figure}
  \centering
  \includegraphics{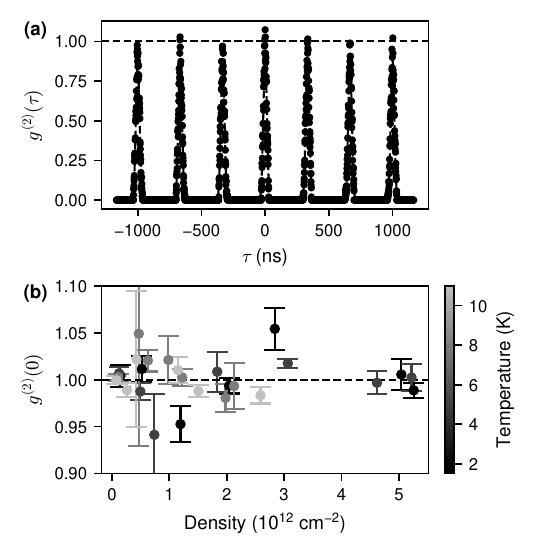}
  \caption{(a) Hanbury Brown and Twiss experiment using pulsed excitation on sample 2. Second-order correlation function $g^{(2)}(\tau)$ for a temperature of 1.7 K and 140 \textmu W. Note the photoluminescence lifetime of the lowest energy peak equals $\approx$ 30 ns (b) Extracted $g^{(2)}(0)$ for varying densities. Total densities are again calculated from PL histograms (cf.\ Fig.~S5). For sample 2, the exciton density saturates at $(5.6 \pm 0.7)\times 10^{12}$~cm$^{-2}$.}
  \label{fig:4}
  \phantomsubcaption\label{fig:4:a}
  \phantomsubcaption\label{fig:4:b}
\end{figure}

Having established the temperature-density diagram, we now investigate how the underlying photon emission statistics evolve with exciton density. We perform a Hanbury Brown and Twiss (HBT) experiment using a pulsed excitation on sample 2. This sample exhibits similar spectral and time-resolved photoluminescence characteristics as sample 1 and a maximum achievable exciton density of $5.6\times 10^{12}$~cm$^{-2}$ (cf.\ Fig.~S5). \cref{fig:4:a} shows a typical trace of the second-order autocorrelation function $g^{(2)}(\tau)$ vs.\ the photon arrival time $\tau$. The repetition frequency of the laser is 3 MHz (i.e.\ 1/3 \textmu s), such that the photoluminescence decays between the laser pulses. The nominal time resolution of the autocorrelation measurement is 25 ps. We observe that $g^{(2)}(0)$ remains close to unity across all investigated densities [cf.\ dashed lines in \cref{fig:4:a,fig:4:b}], suggesting Poissonian photon emission from interlayer excitons throughout the investigated densities and temperatures at the given time-resolution.

\section{Discussion}\label{sec:discussion}

The temporal coherence time in IX ensembles as formed in MoSe$_2$/WSe$_2$ heterostructures is on the order of 400-700fs at low temperature \cite{Sigl2020,Troue2023,Durmus2024}, which is much shorter than the time-resolution of the utilized autocorrelation measurement ($\sim$25 ps, \cref{sec:methods}). Only for a time resolution on the order of the temporal coherence time \cite{Wiersig2009} and in the case of thermal light, one expects a value of $g^{(2)}(0) > 1$. Still, the measured Poissonian behavior is consistent with the exponential decay traces in the time-resolved photoluminescence measurements as in \cref{fig:1:c} and S5. In other words, the photoluminescence via peak (1) seems to be consistent with statistically independent photon emission events from the underlying exciton ensembles \cite{Loudon2000}. We note that we restricted the HBT measurement to sample 2, which we consider representative for both samples given their consistent spectral and time-resolved photoluminescence characteristics.

As far as the temperature-density diagrams are concerned, we observe for both, sample 1 (\cref{fig:1,fig:2,fig:3}) and sample 2 (Figs.~\labelcref{fig:4} and S5) that there is a maximum exciton density for the peak (1) sub-ensemble of $(2.4 \pm 0.1)\times 10^{13}$~cm$^{-2}$ and $(5.6 \pm 0.7)\times 10^{12}$~cm$^{-2}$, which we extract from the time-resolved photoluminescence characteristics of peak (1) [\cref{fig:1:c} and \cref{sec:methods}]. Increasing the laser power further, we make three observations: (i) the spectral width of peak (1) increases (cf.\ Fig.~S7), (ii) the extracted, relative amplitude reduces, while the second peak (2) starts to evolve [\cref{fig:2:a,fig:1:b}], and (iii) the extracted exciton density of peak (1) reduces for the lowest temperatures, while the overall intensity increases [\cref{fig:3:c}]. Generally, the extracted maximum densities are consistent with the ones, where a Mott-transition is predicted for the investigated van-der-Waals heterostructures \cite{Fogler2014}. Looking at the spread of photoluminescence lifetimes at 1.7 K for sample 1 [\cref{fig:1:d}], we tend to neglect laser induced heating effects, because the lifetimes vary less as a function of the investigated range of laser powers than they do when the bath temperature is increased up to 5 K. We note, however, that observations (i), (ii), and (iii) are consistent with, but not uniquely indicative of a Mott transition. One alternative is a phase-space filling of localized moir{\'e}- and/or reconstruction-related states \cite{Richter2024,Figueiredo2025a}. However, filling of individual, quasi-0D moir{\'e} trap sites is unlikely to explain our observations, because such a filling should manifest as inhomogeneously broadened, spiky emission lines already at very low excitation powers, together with a pronounced, discrete blueshift and an asymmetric lineshape as individual sites saturate \cite{BrotonsGisbert2024,Shanks2021,Kremser2020}. None of the mentioned characteristics are observed for peak (1) [\cref{fig:1:b}]. However, a photo-doping-driven renormalization of moir{\'e}- and/or disorder-induced modulations of the potential landscape seems to be plausible on a qualitative level: the photoexcited charge carriers that form the IXs populate local potential minima arising, e.g., from disorder, reconstruction, and weak moir{\'e} effects. In this way, an increasing number of electron-hole pairs renormalizes the potential landscape via Coulomb-interactions such that the overall potential is flattened out. Tentatively, this scenario explains the decrease of peak (1) accompanied with an increase of peak (2) [\cref{fig:2:a,fig:3:d}] as well as the approximately fourfold difference in saturation density between the two samples. In this interpretation, peak (1) resembles a sub-ensemble of IXs, where the center-of-mass wavefunction can be extended across several lattice sites \cite{Richter2024,Figueiredo2025a}, while peak (2) stems from the photon emission of shorter lived IXs occupying a dynamically flattened potential landscape. In turn, we argue that the observations (i), (ii), and (iii) are consistent with a combination of a photodoping-induced renormalization of moir{\'e}- and disorder-induced modulations of the potential landscape, and/or a possible contribution from a Mott transition of the investigated IX ensemble underlying peak (1).

The observed Poissonian photon emission statistics is consistent with an ensemble of excitons acting as classical particles \cite{Loudon2000} as well as with a possible bosonic many-body state \cite{Fogler2014,Sigl2020,Troue2023,Katzer2023,Cutshall2025}. We find no significant deviation of $g^{(2)}(0) = 1$ for high excitation powers, i.e.\ where the excitonic picture might not apply because of a possible Mott-transition, and for bath temperatures, where a transition to a phase of an ideal gas is expected \cite{Fogler2014,Troue2023}. In the latter case of a possible thermal photon emission, we seem to be limited by the time resolution of the autocorrelation measurement \cite{Wiersig2009}. Corresponding measurements with a sub-picosecond time resolution, however, are beyond the scope of the current study.

\section{Conclusion}\label{sec:conclusion}

This work presents a density-temperature diagram of interlayer excitons in MoSe$_2$/WSe$_2$ heterostructures, combining time-resolved photoluminescence with second-order correlation measurements. We observe near-unity second-order correlations ($g^{(2)}(0) \approx 1$) from low exciton densities up to the maximum achievable density, which demonstrates that the interlayer exciton ensemble maintains Poissonian emission statistics. Both investigated samples exhibit a saturation of the luminescence peak that is assigned to the ensembles with the lowest energy. A careful, combinatorial analysis of the photoluminescence lifetime, emission intensity, and propagation characteristics reveals a threshold exciton density of order $10^{13}$~cm$^{-2}$, which is consistent with a possible Mott transition. An alternative interpretation is given by photo-doping induced renormalization effects flattening the exciton potential landscape.

\section{Methods}\label{sec:methods}

\subsection{Sample preparation}\label{sec:methods:sample}

The investigated heterostructures consist of micromechanically exfoliated monolayers of MoSe$_2$ and WSe$_2$ stacked with a twist angle close to 60$^\circ$ (nearly H-type stacking, with a twist-angle accuracy of $\pm 0.5^\circ$). After exfoliation, we use viscoelastic stamping to encapsulate the monolayers in h-BN and position them on a Si/SiO$_2$-substrate. The heterostacks are vacuum-annealed at 400 K for two hours.

\subsection{Photoluminescence measurements}\label{sec:methods:pl}

The photoluminescence signal is collected with a single-mode fiber and analyzed in a spectrometer. The integration times are logarithmically scaled from 1000 s at the lowest to 25 s for the highest investigated power.

\subsection{Time-resolved measurements}\label{sec:methods:trpl}

The time-resolved measurements are performed using an avalanche photodiode in combination with a time-correlated single photon counting module. The base resolution of the TCSPC module amounts to 25 ps with a timing precision of $<$20 ps RMS. By using an additional filter wheel in the detection path, we limit the count rate of the photodiode to $2\times 10^{5}$~s$^{-1}$, so that the time between photon detections stays below the dead time of the photodiode. The measurements are done with a fixed integration time of 20 s.

\subsection{Second-order correlation measurements}\label{sec:methods:g2}

The interlayer exciton emission is detected after a 50/50 single-mode fiber splitter with two avalanche photodiodes. Transistor--transistor logic pulses are recorded by a time-correlated single photon counting module in a T2 measurement, where every photon event gets recorded. The base resolution of the TCSPC module amounts to 25 ps with a timing precision of $<$20 ps RMS. The laser's gating signal is recorded in an additional channel. This allows cross-correlation over arbitrary time lags after the measurement relative to the timing of the laser pulse. The second-order autocorrelation function at zero time delay $g^{(2)}(0)$ is calculated as the ratio of the integrated coincidence counts at zero time delay (central peak in \cref{fig:4:a}) to the average coincidence counts at non-zero time delays.

\subsection{Fitting of decay histograms}\label{sec:methods:fit}

Photoluminescence time traces $I(t)$ are modeled using exponential decays of the form $I_0 e^{-t/\tau}$. Additionally, the instrument response function (IRF), with the convoluted avalanche photodiode's timing jitter and the laser turnoff time, which amounts to 1.55 ns, is included in the fit function. Moreover, we assume the lifetime of individual exciton species to be time-independent. Time-dependent effects, such as cooling, would show up as an additional decay channel with a characteristic lifetime. A two-exponential decay, including the system's IRF, yields a good description of the time traces as shown in Figs.~S2 and S3 and matches the observation of separated PL peaks in spectroscopic measurements as already previously demonstrated \cite{Figueiredo2025b}.

\subsection{Density estimation}\label{sec:methods:density}

We note that the calculated densities consider only decay processes and thus provide a lower bound on the number of excitons. The exciton loss from the collection spot at elevated excitation powers introduces another loss channel for excitons in the single-mode measurement. The local amplitude is measured with a single-mode fiber with a mode field diameter of 5\textmu m ($A_\mathrm{SM} = 0.48$~\textmu m$^2$) and the global lifetime is measured with a multi-mode fiber with a core size of 105\textmu m ($A_\mathrm{MM} = 210.2$~\textmu m$^2$). The reported densities are calculated according to
\begin{equation}
  n = \frac{I\tau}{A\eta}\quad,
  \label{eq:1}
\end{equation}
using the collection efficiency $\eta = 0.174$ (cf.\ table S1), which includes the emission characteristics of the optical dipole within our specific heterostructures, all components in the optical path, free-space coupling into the fiber and the quantum efficiency of the avalanche photodiode \cite{Sigl2022}.

\begin{acknowledgments}
The authors gratefully acknowledge the German Science Foundation (DFG) for financial support via Grants HO 3324/16-1, No.~290642686, and 556436549 (WU637 4-2, 8-1), and the clusters of excellence MCQST (EXS-2111) and e-conversion (EXS-2089), and the priority program 2244 (2DMP) via HO3324/13-2 and WU637 7-2 (No.~443274199). K.W. and T.T. acknowledge support from the JSPS KAKENHI (Grant Numbers 21H05233 and 23H02052), the CREST (JPMJCR24A5), JST and World Premier International Research Center Initiative (WPI), MEXT, Japan.
\end{acknowledgments}

\bibliography{mott}

\end{document}